\documentclass[reprint,superscriptaddress,singlecolumn]{revtex4}
\usepackage{graphicx}
\usepackage{epsf}
\usepackage{caption}
\usepackage{subfig}

\begin{document}

\newcommand{\newc}{\newcommand}
\newc{\mbf}{\mathbf}
\newc{\boma}{\boldmath}
\newc{\beq}{\begin{equation}}
\newc{\eeq}{\end{equation}}
\newc{\beqar}{\begin{eqnarray}}
\newc{\eeqar}{\end{eqnarray}}
\newc{\beqa}{\begin{eqnarray*}}
\newc{\eeqa}{\end{eqnarray*}}
\newc{\bd}{\begin{displaymath}}
\newc{\ed}{\end{displaymath}}

\title{Generalized ballistic deposition in 2 dimensions : \\ scaling of surface width, porosity and conductivity}
\author{Subhankar Ray}
\email{sray@phys.jdvu.ac.in}
\affiliation{Department of Physics, Jadavpur University, Calcutta 700 032, India.}
\author{Baisakhi Mal} 
\email{baisakhi.mal@gmail.com}
\affiliation{Budge Budge Institute of Technology, Calcutta 700 137, India}
\affiliation{Department of Physics, Jadavpur University, Calcutta 700 032, India.}
\author{J. Shamanna}
\email{jlsphy@caluniv.ac.in}
\affiliation{Physics Department, University of Calcutta, Calcutta 700 009, India.}

\begin{abstract}
A deposition process with particles having realistic intermediate stickiness is studied in $2+1$ dimensions.
At each stage of the deposition process, for any given configuration, a newly depositing particle gives rise to
allowed set of configurations that are
vastly larger than those for deposition of a mixture of purely non-sticky (random like) and purely sticky (ballistic like) particles.
We obtain scaling behaviour and demonstrate collapse of scaled data for surface width and porosity.
Scaling of conductivity, when a porous structure thus formed, is saturated with conductive fluid, e.g. brine, is studied.
The results obtained are in good agreement with Archie's law for porous sedimentary rocks.
\end{abstract}
\date{03 March 2015}
\maketitle

\section{Introduction}
The growth of surfaces in different dimensions and on different substrate geometries, finds applications in several areas 
of science and technology, including physics, chemistry, biology, geology,
chemical engineering and material science.
Though the applications are diverse, 
physical properties of growth processes, such as, nature of roughness,
porosity and their dynamic scaling behavior depend on few basic entities, such as, dimensionality,
geometry and underlying symmetry of the problem.
Hence, these systems can be classified into a few universality classes. 
Extensive theoretical and experimental study have been undertaken in these areas.
The theoretical study of dynamic scaling behaviour of surfaces
generally follows two pathways, 
extensive simulations of discrete models \cite{bara95,mea93,fam85,fam86}, and
study of relevant stochastic differential equations obtained
from phenomenological consideration \cite{ew82,kpz89}.
Another relatively recent approach, involves obtaining difference
equations from microscopic deposition rules for the discrete models
and deriving relevant stochastic differential equations by limiting
process using various regularization techniques.
\cite{vve03a,vve03b,hasel07a, hasel07b}.

The simplest growth process is studied by considering a single species of particles,
either completely non-sticky or completely sticky, descending on a one-dimensional
or a two dimensional substrate. These are called random deposition (RD) and ballistic
deposition (BD) respectively.
For more realistic systems, one needs to consider the possibility that,
a single type of particle may have an intermediate stickiness and 
in one deposition process, several such species may be involved.

In order to study intermediate stickiness, Wang, Cerdiera et.al. have studied models with two types of particles,
some random-like, and others ballistic-like\cite{wang93,wang95,nashar00,nashar96}. 
However neither species are allowed to have intermediate stickiness.
Horowitz, Albano \cite{horo06, horo01b} have studied growth models in which each incoming
particle may behave either as non-sticky with probability $p$ or as
completely sticky with probability $(1-p)$.
However, neither mimics the possibility that a particle may be partially sticky.
Thus at each contact with the surface, it may have a fixed probability of sticking ($0 < p < 1$),
and a fixed probability ($q=(1-p) < 1$) of continuing its journey till it settles somewhere on the surface.
Study of such model in ($1+1$) dimension was proposed in an earlier work\cite{bai1dGBD}.
In the present article, we extend the study to ($2+1$) dimensions.
A related model on a one-dimensional substrate , with next nearest neighbor sticking
was studied by Banerjee et al. \cite{kas14}.

In this physically realistic model, the incoming particle
may come in contact with several points on the surface, and its final position of
deposition forms a vastly larger ensemble than that considered in the former
studies \cite{wang93,wang95,nashar00,nashar96, horo06, horo01b}.
In case of (2+1)-dimension, i.e., for growth on a two dimensional substrate,
the present model may find application in the formation of sedimentary rocks and fabrication of nano materials.

The structure of pores in depository rocks is of great importance in rock
geology (petrology) and oil exploration.
Surface roughness is measured in terms of the standard deviation, which is the square
root of the second central moment of the height distribution and is defined as,
\beq\label{surfwid}
W(L,t) = \sqrt{ \frac{1}{L^2} \sum_{i,j=1}^{L} \left[ h(i,j,t) -
\langle h(t) \rangle \right]^2 } ,
\eeq
where, $h(i,j,t)$ is the height of the $(i,j)$-th site at
any instant $t$, $L$ is the system size and $\langle \rangle$ is the average.
Though this is an important quantity of interest, it is far less
informative than the distribution of height itself.
Knowledge of frequency or probability distribution of height is equivalent
to the knowledge of all moments \cite{bai11}.
Similarly, the porosity gives an average information about the nature of the
pores. It cannot, however, tell us if the pores are clumped together, or are more scattered
throughout the allowed volume.
Moreover, it cannot distinguish between situations where pore clusters
are elongated longitudinally or transversely.
Additional information about conductivity may throw some light
on the above mentioned geometry, though in a somewhat qualitative manner.
For a given porosity, higher conductivity implies more clustering
of pores, predominantly in the longitudinal direction, rather than in
circuitous paths having long horizontal parts.
In addition, in real life experiments in geology, this conductivity is more
easily measured than other measures of pore geometry.
Thus the study of conductivity, and its scaling with porosity is important
for better understanding of the porous structure.
In this article we obtain surface width, porosity and conductivity, and their
scaling behavior in (2+1)-dimensional growth on a flat substrate.

\begin{figure}[!ht]
\center
  \subfloat[RD]{\includegraphics[width=0.25\linewidth]{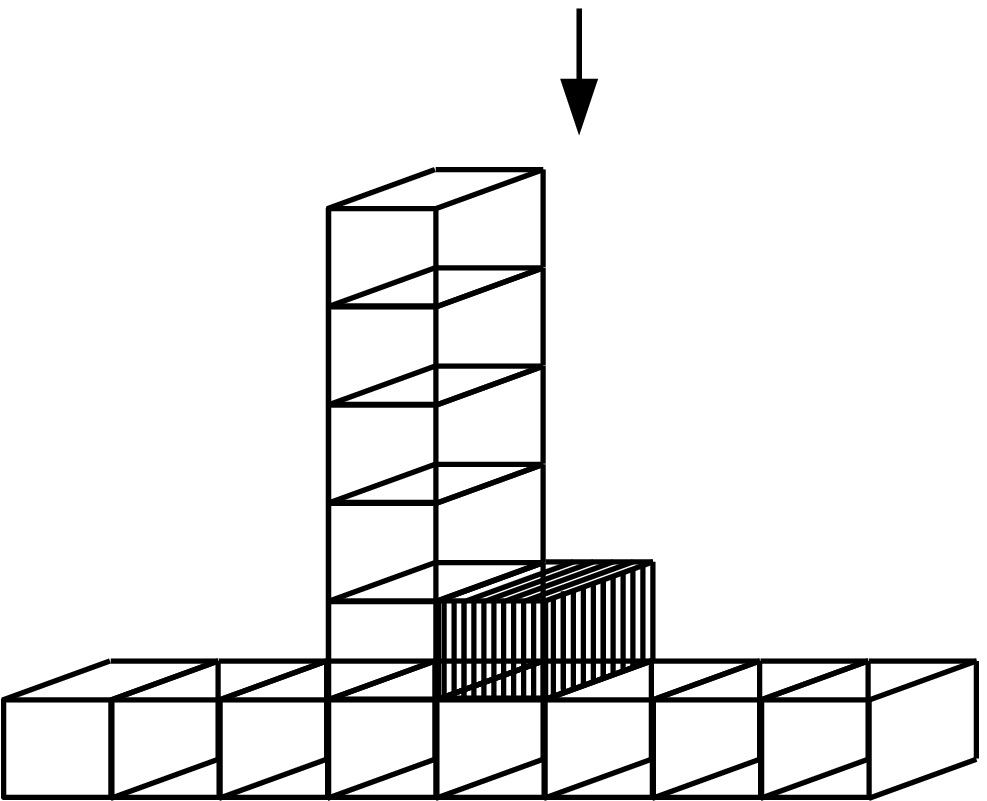}
  \label{RD}}%
  \qquad
  \subfloat[BD]{\includegraphics[width=0.25\linewidth]{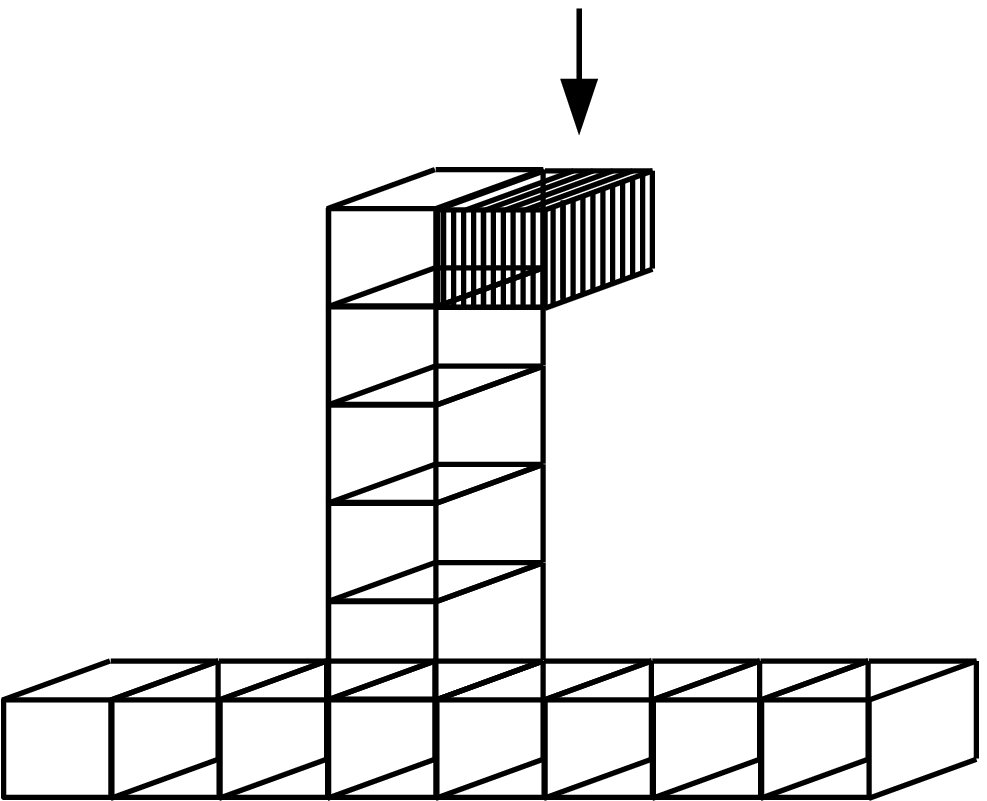}
  \label{BD}}%
  \qquad
  \subfloat[GBD]{\includegraphics[width=0.25\linewidth]{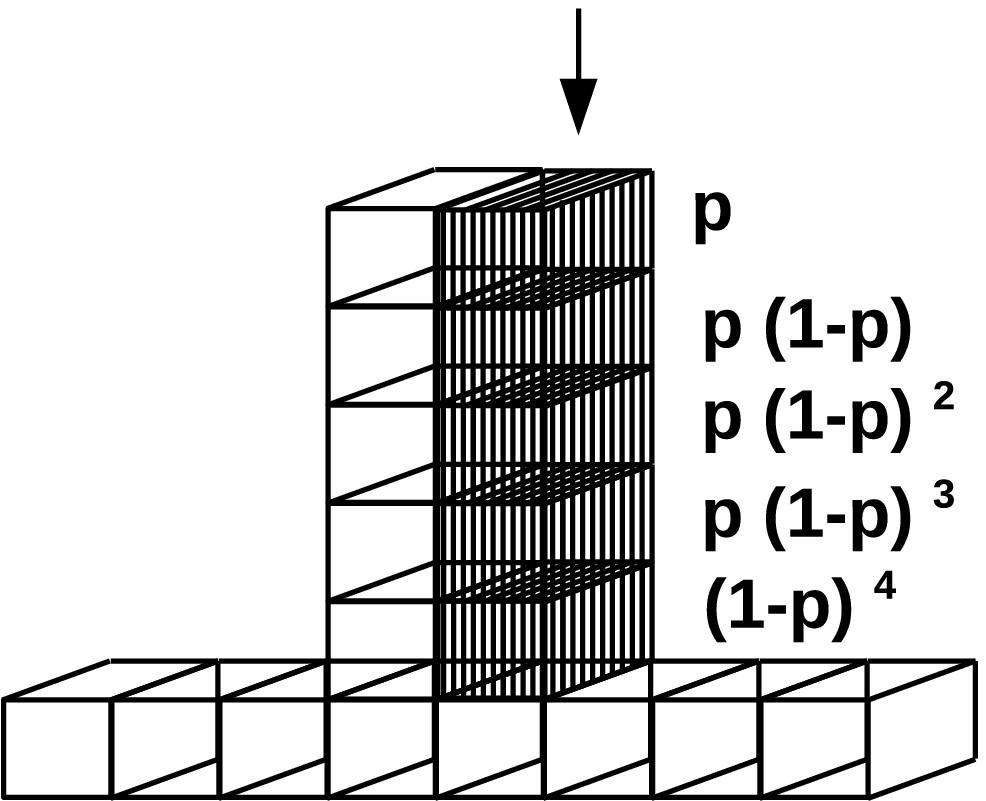}
  \label{gRDBD}}
\caption{Allowed positions in (a) RD (b) BD (c) GBD 
\label{genRDBD}}
\end{figure}

In RD the individual columns grow independently of each other
without any bound and thus roughness of the interface width
never saturates. No voids are present within the composite
\cite{bara95}.
Correlations can be introduced by making the particles sticky.
The model representing such a system is called ballistic deposition
(BD) model, where an incoming particle sticks to the first point
of contact it encounters with the surface while falling down vertically
towards a randomly chosen site on the substrate \cite{bara95, fam85,fam90}.
The stickiness in BD is extreme, the particle must stick at the very first
contact and is given no option to slide past the first point of contact.
In real systems, one may find a particle sticking to a site after sliding
past a few points of contact on its vertical journey.
In this article we extend our earlier work \cite{bai1dGBD} of such a realistic process,
to surfaces growing on a two-dimensional substrate
and building a three-dimensional structure with voids.
This porous structure may fill up with a conducting fluid, such as,
brine, as in the case of sedimentary rocks.
The conductivity of the structure depends on the conductivity of the brine, 
the relative amount of pores, and the geometry of the pores in the 
3-d structure.
To begin with, we take the specific conductivity of brine as a constant.
The porosity and geometry of the pores both depend on the 
stickiness and possibly the size of the substrate.
The variable stickiness of particles is modeled using a parameter $1\ge p\ge 0$.
A particle dropped on to the substrate sticks to the first surface it
encounters with a probability $p$ and continues on its downward journey
with probability $(1-p)$. The probability that it will deposit
at the very next surface it encounters is $p(1-p)$. It will continue to
the next lower position with probability $(1-p)(1-p)$. Thus, if a site is
selected with taller nearest neighbor, a newly arriving particle
can deposit at any one of the successive positions $1,2,3 \dots$ shown in
Fig. \ref{gRDBD}, with probabilities $p, p(1-p), p(1-p)^2 \dots$
respectively. In this model, $p=0$ corresponds to RD (Fig. \ref{RD})
and $p=1$ corresponds to BD (Fig. \ref{BD}), while $0<p<1$ represents
intermediate stickiness.
It may be noted that the present model does not allow sticking at corners or on edges,
that is it disallows sticking to next nearest neighbors. 
In the present model the surface width depends on the sticking 
probability $p$ and system size $L$,
in addition to time $t$, measured in terms of the average number of layers deposited
during that time.
The logarithmic plot of surface width $W$ versus time $t$ shows four distinct regions. 
There is an initial random like growth region (GR-1),
followed by a non-KPZ like growth (GR-2) and then a KPZ growth (GR-3) followed
by an eventual saturation $W(L,p,t) \rightarrow W_{sat} (L,p)$.
Similar feature was also reported earlier in one dimension \cite{bai1dGBD}, 
and in systems where particles may stick to next nearest neighbors \cite{kas14}.
With the introduction of probability of sticking $p$, the Family-Vicsek
scaling relation $W(L,t) \sim L^{\alpha} f(t/L^z)$ is modified and the following
dynamic scaling relation obtained for the growth and saturation regions,
\beq\label{wscale}
W(L,p,t) \sim L^{\alpha}p^{-\alpha'}F\left(\frac{t \, p^{z'}}{L^z}\right),
\eeq
where $F(x)$ is scaling function that satisfies $F(\infty) \sim constant$
and $F(x) \sim x^{\beta}$ for small $x$.
The scaling exponents are determined and an excellent collapse of scaled
data is obtained using those exponents.

We also study the porosity of the bulk of size $L^3$ just beneath the 
active region, which is signified by the volume inaccessible to the new
incoming particles. It is found that the porosity reaches a
constant value almost as soon as a bulk volume of dimension $L^3$
forms below the active region. 
We find a scaling relation between saturated porosity
$\rho_s$, system size $L$ and probability of sticking $p$ as,
\beq
\rho_s \sim p^a L^b.
\eeq
The porous structures thereby formed are further investigated
for their conducting properties. The saturated conductivity $\sigma_s$
is found to obey a scaling relation
\beq
\sigma_s \sim p^m L^n.
\eeq
The corresponding exponents are calculated from our simulational results.
In addition we observe that $(\sigma_s/L)$ depends on $\rho_s$
as $(\sigma_s/L) \sim \rho_s^f$ with $f = 2.02$, which is
in good agreement with Archie's law,
an important empirical law in geophysics \cite{archie1,archie2,archie001}.

In our simulations of the present generalized deposition model,
two independent random number generators were
used, one for selecting a site on the growing surface and,
another to determine whether a particle will stick
at a particular location for a chosen value of the sticking probability.
These two random number generators are chosen to ensure that they are
completely independent and uncorrelated.
The reliability of the random numbers used is verified by the $\chi^2$-test
and absence of repetitive subsequences or looping for the maximal
set of random numbers drawn for both the sets.


\section[Generalized Ballistic Deposition (GBD)]{Generalized Ballistic Deposition (GBD) - Variable stickiness}
The present model is a modification of ballistic 
deposition to represent
realistic sticky particles. The model is studied in (2+1)-dimensions. 
A particle is allowed to descend vertically towards a randomly chosen site on a
two dimensional substrate. If the selected site is higher than its nearest neighbors,
the particle simply deposits on top of the column at that site.
However, if the chosen site has a taller column of particles as its nearest neighbor, then 
the new particle sticks to the first occupied site it encounters if
the value of $p$ is larger than a random number generated 
from a uniform distribution between 0 and 1. Otherwise, it slides
down vertically to the next occupied site with probability $(1-p)$.
At this site the particle may stick with probability $p(1-p)$
or continue its further descent with probability $(1-p)^2$, and so on, till it
reaches the bottom.
Thus if the chosen site has a nearest neighbor with column height taller
by $n$ layers relative to it, the probabilities of the arriving particle sticking to the
successive particles of the nearest neighbor column from top are given by,
\beq
P(1) = p, \; P(2) = p(1-p), \dots \; P(k) = p(1-p)^{(k-1)}.
\eeq
The probability that the particle slides past the preceding $(n-1)$
occupied neighbors, and lands at the lowest possible position is given by,
\bd
P(n) = 1 - \sum_{k=1}^{n-1} P(k) = (1 - p)^{(n-1)}.
\ed
This describes a proper stochastic process. The total probability of a descending particle sticking
to one of the allowed position is 
$\sum_{k = 1}^{n} P(k) = 1$.
It must be noted that sticking to corners or edges are not allowed. 
Only surface sticking is allowed in this particular model. 

\section{Results and discussions}

\subsection{Scaling of surface width}
\begin{figure}[!h]
\begin{minipage}{0.49\linewidth}
  \includegraphics[width=\linewidth]{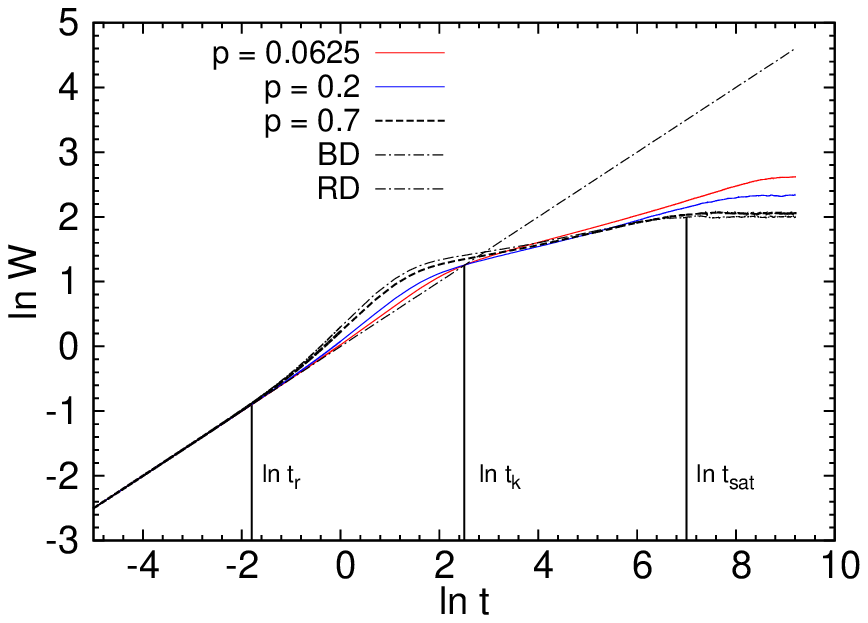}
  \caption{Logarithmic plot of interface width 
and time for different $p$ for $L = 256$. \label{lnwvslnt_L}}
\end{minipage}
\begin{minipage}{0.49\linewidth}
 {\includegraphics[width=\linewidth]{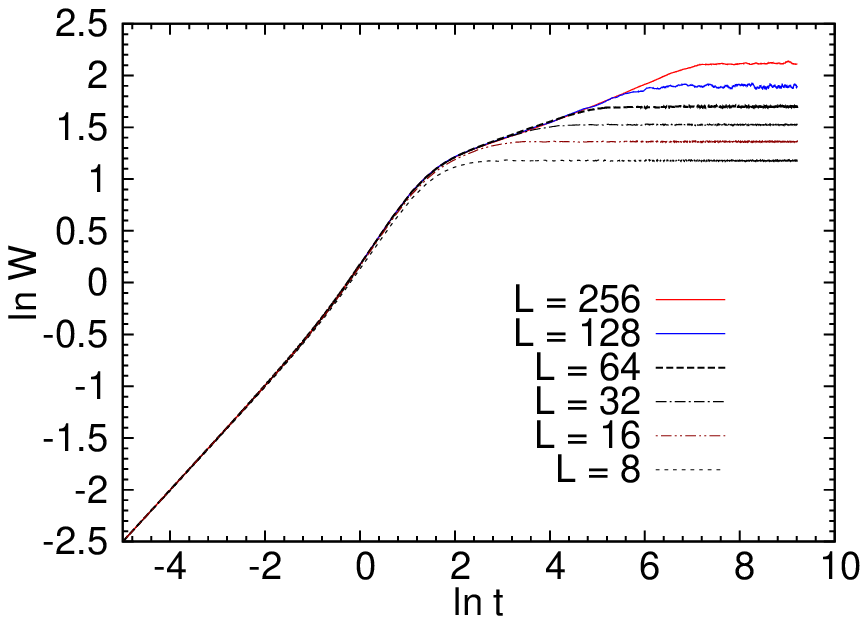}}
  \caption{Characteristic plot for $\ln W_{sat}$ vs $\ln t$ with $p=0.5$
for different system sizes.  \label{lnwvslnt_p}}
\end{minipage}
\end{figure}

Simulations have been performed for several system sizes
and various probabilities of sticking.
We present the analysis of data for system sizes
$L = 64, 128, 256, 512$ and
values of probability of sticking
$p = 1, 0.8, 0.7, 0.5, 0.25, 0.125, 0.0625$.
The $p=0$ gives us the random limit and $p=1$ is the ballistic deposition.
The logarithmic plot for surface width and time shows
four distinct regions with varying slopes as shown in Fig. \ref{lnwvslnt_L}.
It is interesting to note that this feature, namely the existence
of four characteristic regions, is observed whenever
stickiness is present, i.e. for all non-zero probability of sticking $p$,
however small.

The dependence of surface width $W$ on $t$ in log-log scale,
in the early submonolayer region ($t \ll 1$) is linear 
with slope $1/2$ as in random deposition (growth region 1, GR-1). 
At later stages of submonolayer growth (growth region 2, GR-2), $t \simeq 1^-$, the
surface width shows a steep increase which continues for the 
first few layers ($1-\epsilon \le t\le 3$, $ 1 \gg \epsilon > 0$). With deposition of further layers, 
the rate of increase in width slows down (growth region 3, GR-3). 
After deposition of a large number of layers, the ensemble average of the surface width 
saturates. Three different crossover times are of 
relevance. The first crossover time $t_r$ corresponds to the change 
from random growth to region with slope greater than $1/2$. 
The second crossover time $t_k$ corresponds to time beyond few layers 
where the slope decreases and changes from GR-2 to GR-3. 
The third crossover time $t_{sat}$ corresponds to beginning 
of saturation region.

\begin{figure}[!htb]
\begin{minipage}{0.49\linewidth}
\center
  \includegraphics[width=\textwidth]{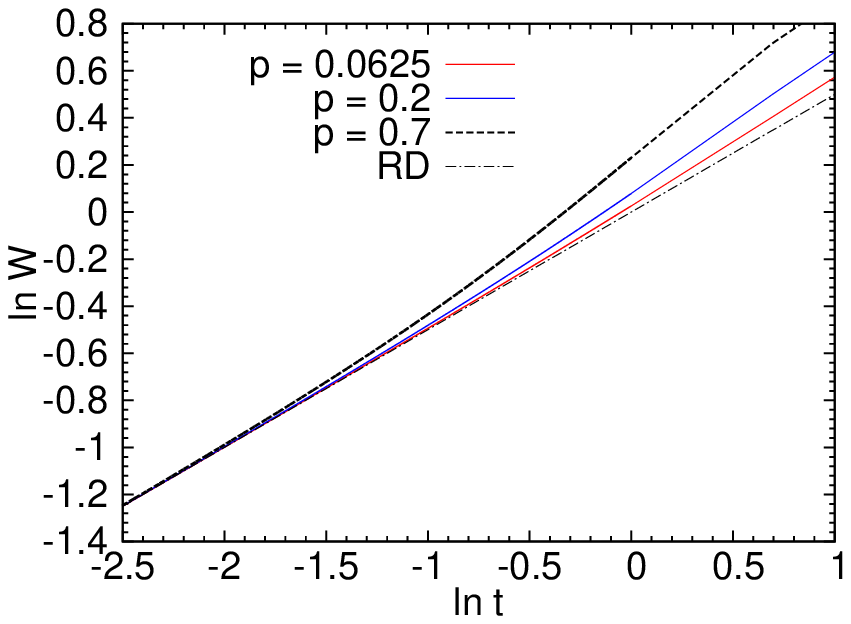}
  \caption{Deviation from the random like behavior at later stages of
    submonolayer regime for $L = 256$.} 
  \label{devrandom}
\end{minipage}
\begin{minipage}{0.49\linewidth}
  {\includegraphics[width=\textwidth]{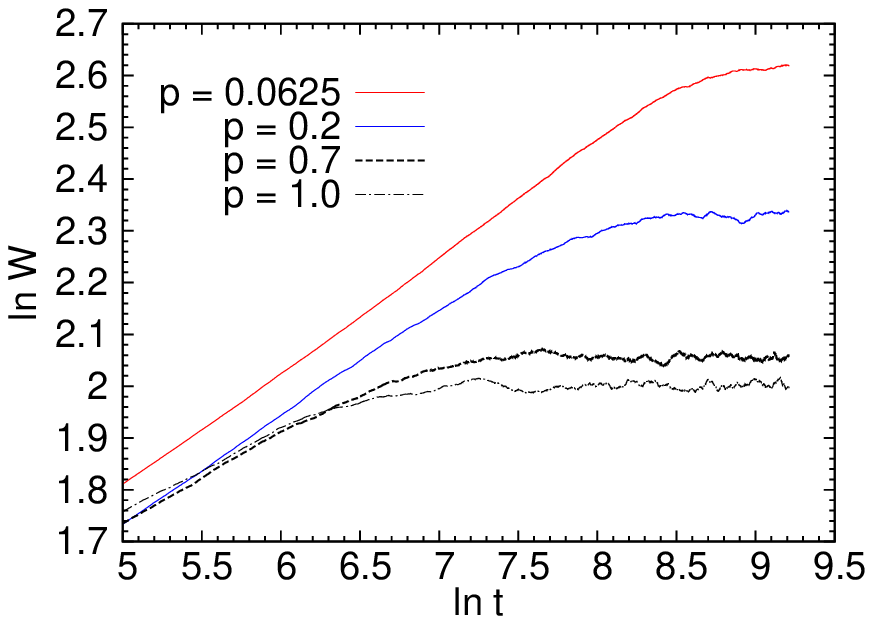}}
  \caption{Dependence of width in later stages of deposition with saturation for 
$L = 256$.}
  \label{satplot}
\end{minipage} 
\end{figure}

The appearance of different growth regions in the present
model may be understood as follows.
In our model, we start from a flat substrate, thus initially almost no two adjacent
sites are occupied, hence there is no correlation among neighboring columns.
Thus for system of all sizes, at the very beginning, when $t \ll 1$, the growth is random like,
irrespective of whether we allow sticking or not 
(Fig. \ref{lnwvslnt_L} and Fig. \ref{lnwvslnt_p}).
The deviation from random like behavior begins near the first monolayer,
and may continue for few monolayers of deposition.
In this region the surface width grows at a rate faster than
that in the case of random deposition.
As the number of particles deposited at this stage is nearly $L$,
due to fluctuation, some short multi-layer columns begin to form.
This brings in non-trivial correlations in the system,
due to possibility of the descending particles encountering occupied neighbors
before reaching the bottom of their own columns.
At this stage, the average height of the surface is small,
and even a few particles sticking to a higher location instead of reaching
the bottom of a column, makes a significant relative change in width.
Thus the rate of growth of surface width in this region is higher than
that for RD (Fig. \ref{devrandom}). 

With further layers of particles being deposited, we reach a third growth region GR-3,
where the rate of increase in surface width slows down (Fig. \ref{satplot}).
The average height and the interface width are large in this region.
The deep crevices are encountered by descending particles, and if they are sticky,
they can stick to a side wall, thus filling up the crevices much faster
than for RD, where the crevices need to be filled from bottom up.
At an even later time, the above mentioned smoothening effect starts dominating,
and the surface width finally saturates.
The saturated width depends both on the system size $L$ and sticking probability $p$.

\begin{figure}[!htb]
\center
\begin{minipage}{0.49\linewidth}
  {\includegraphics[width=\textwidth]{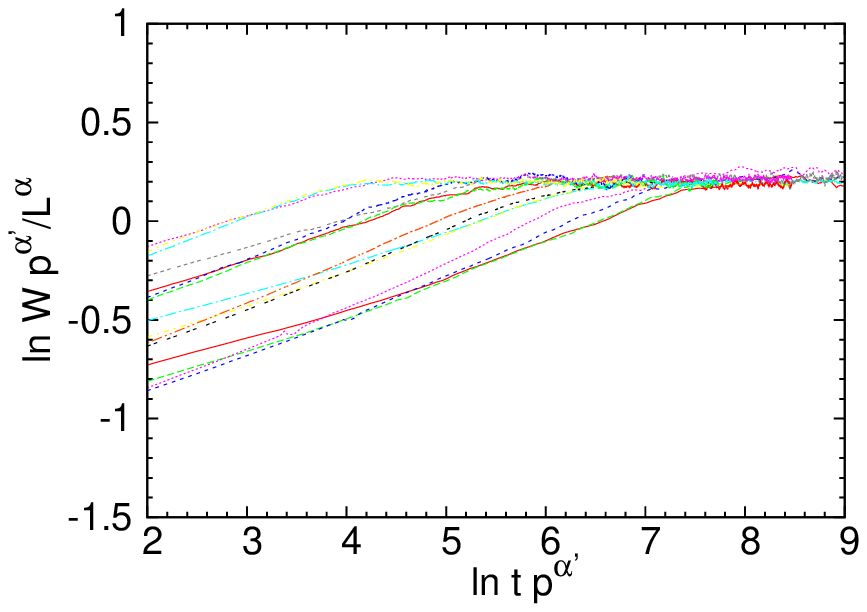}}
  \caption{ $\ln W p^{\alpha^{\prime}}/L^{\alpha} $ versus  $\ln t p^{z^{\prime}}$ \label{scale03}}
\end{minipage} %
\begin{minipage}{0.49\linewidth}
  {\includegraphics[width=\textwidth]{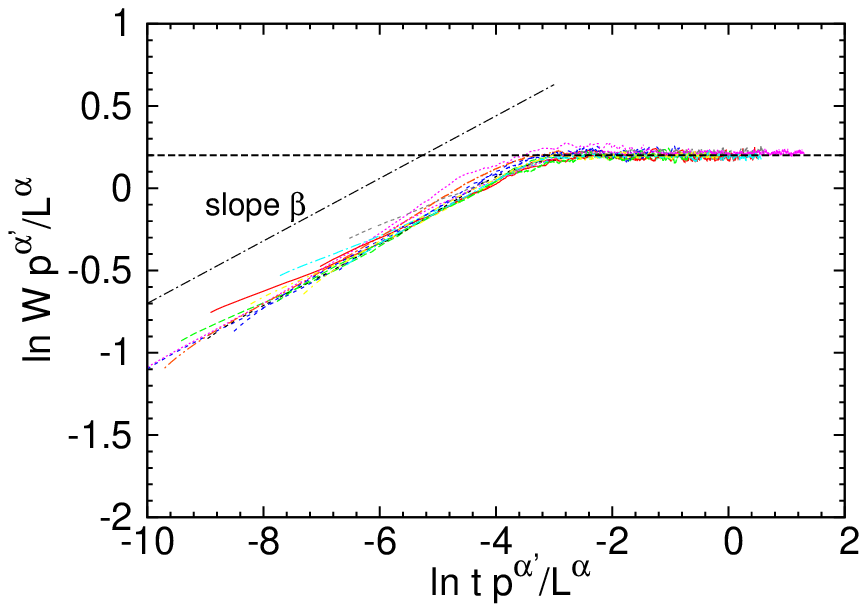}}
  \caption{$\ln W p^{\alpha^{\prime}}/L^{\alpha}$ versus  
$\ln t p^{z^{\prime}}/L^z$. \label{fscale}}
\end{minipage} 
\end{figure}

For a given value of $p$, the saturated width $W_{sat}$ and $t_{sat}$
increase with system size $L$ and for fixed $L$, they
decrease with increase in 
probability of sticking $p$. This decrease is more pronounced for lower 
values of stickiness parameter, i.e., $p\le 0.5$. 
The KPZ-like growth region GR-3, and
the saturation region follow scaling relation stated in Eq.\ref{wscale},
for which the exponents can be evaluated.
With the increase in the probability of sticking $p$, the saturation is
at lower values of interface width. This dependence is found to be 
of the form $p^{-\alpha^{\prime}}$ with $\alpha^{\prime} = 0.20846$.
\begin{figure}[!htb]
\center
\begin{minipage}{0.49\linewidth}
  {\includegraphics[width=\textwidth]{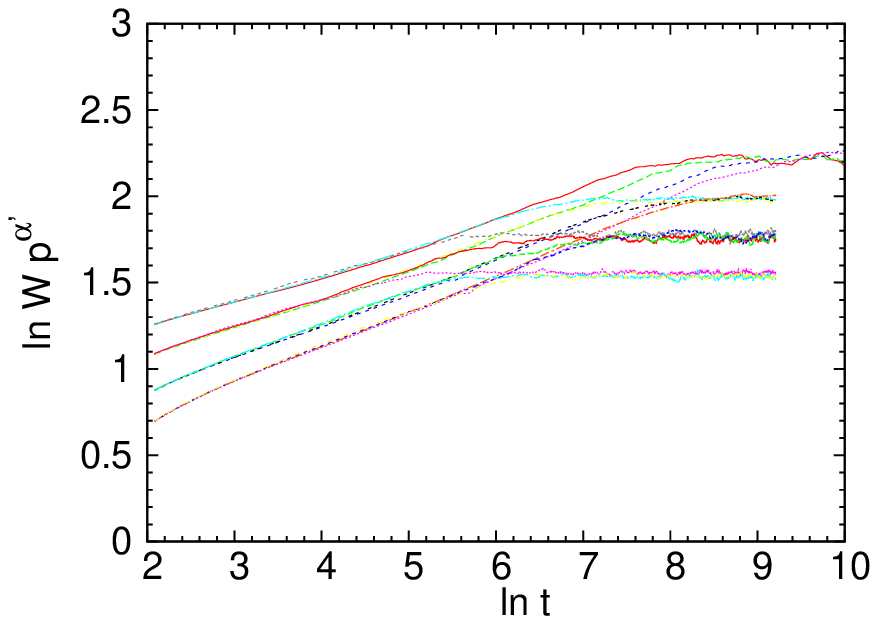}}
  \caption{$\ln W p^{\alpha^{\prime}}$ versus  $\ln t$  \label{scale01}}
\end{minipage}%
\begin{minipage}{0.49\linewidth}
 {\includegraphics[width=\textwidth]{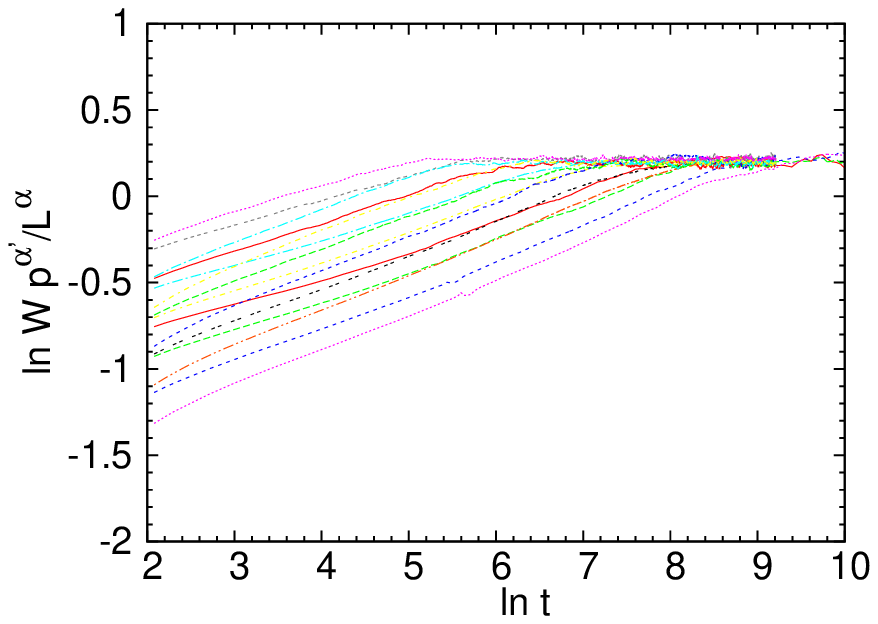}}
  \caption{$\ln W p^{\alpha^{\prime}}/L^{\alpha}$ versus  $\ln t$    \label{scale02}}
\end{minipage} 
\end{figure}
Further, this partially scaled width $\ln W \, p^{\alpha^{\prime}}$ 
depends on the system size as $L^{\alpha}$
with $\alpha = 0.3224$ (Fig. \ref{scale01}). The plot of scaled width 
$(\ln W \, p^{\alpha^{\prime}}/L^{\alpha})$
versus $\ln t$ is shown 
in Fig. \ref{scale02} for $ p = 0.8, 0.5, 0.25, 0.125$ for system sizes
$L = 64, 128, 256, 512$. The figure shows collapse of scaled data in the saturation region.
We observe that the KPZ-like growth region is more pronounced, and occurs
over a wider time interval for larger system sizes, 
as the correlation effects take a longer time to bring in the saturation.
However, with increasing probability of sticking $p$, the correlation effects
are more dominant and hence saturation kicks in much earlier, thereby
resulting in a shorter KPZ-like growth region.
The scaling for the growth region with respect to sticking probability is shown
in Fig. \ref{scale03}, and the complete scaling is shown in Fig. \ref{fscale},
using the our calculated values for exponents $z = 1.72348$ and $z^{\prime} = 1.069246$.
The slope of the fully scaled width versus time, in the growth region GR-3 is
determined as $\beta=.19$. 
Thus in growth region we get,
\beq\label{wscale02}
W(L,p,t) \sim L^{\alpha}p^{-\alpha'}F\left(\frac{t \, p^{z'}}{L^z}\right) \rightarrow 
L^{\alpha}p^{-\alpha'} \left(\frac{t \, p^{z'}}{L^z}\right)^\beta, \;\; \text{for small} \;
\left(\frac{t \, p^{z'}}{L^z}\right).
\eeq
Whence we identify the following relation connecting the obtained scaling exponents,
\beq\label{scalingrel}
\beta = \frac{\alpha}{z} = \frac{\alpha^\prime}{z^\prime},
\eeq
where $\alpha = .3224 \simeq 1/3$, $z = 1.7234 \simeq 5/3$, $\alpha^\prime = .20846 \simeq 1/5$,
$z^\prime = 1.069 \simeq 1$ and $\beta = .19 \simeq 1/5$.

\subsection{Porosity and Conductivity}
\begin{figure}[!htb]
\begin{minipage}{0.49\linewidth}
\center
  {\includegraphics[width=\linewidth]{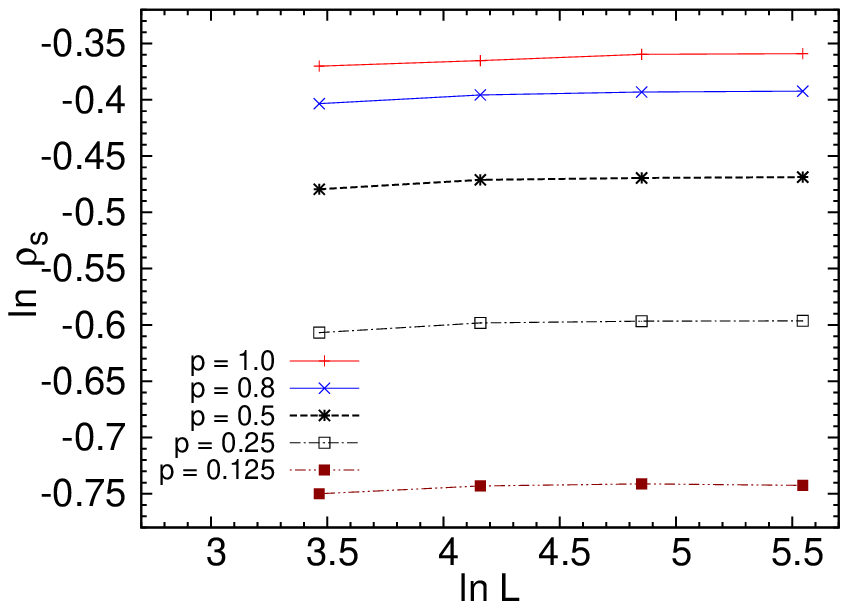}}
  \caption{Variation of $\rho_s$ with system size $L$ in log-log
scale for different values of $p$. \label{por01}}
\end{minipage} %
\begin{minipage}{0.49\linewidth}
\center
  {\includegraphics[width=\linewidth]{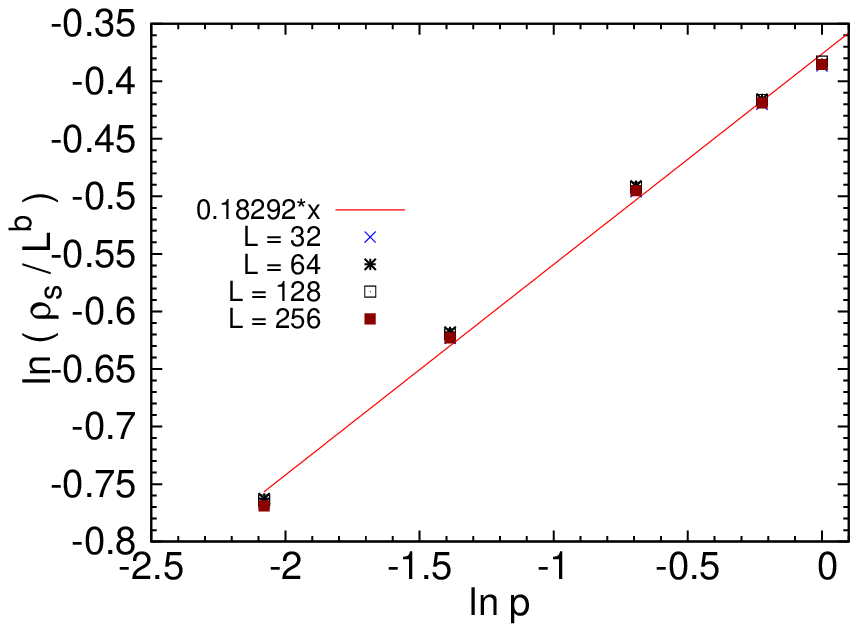}}
  \caption{$\ln \rho_s /L^b$ versus $\ln p$. \label{por03}}
\end{minipage} 
\end{figure}

In this section, we study the dependence of porosity on system size $L$. 
Porosity $\rho$ is defined as the number of vacant sites within a 
cubic volume of side $L$, just beneath the active surface of the deposit.
For $(2+1)$-dimensional systems, it is found that the porosity reaches 
a saturation value by the time a volume $L^3$ is formed below the active layer.
Thus, there is no measurable variation of the porosity with time.
The dependence of the porosity on the system size $L$ is also not very
significant (Fig. \ref{por01}).
The deposit becomes more porous for higher sticking probability $p\gg0$.
The saturated porosity $\rho_s$ scales with $p$ and $L$ as,
$\rho_s \sim (p^a L^b)$ with $a = 0.182920$ and $b = 0.00478$.

\begin{figure}[!htb]
\begin{minipage}{0.49\linewidth}
\center
  {\includegraphics[width=\linewidth]{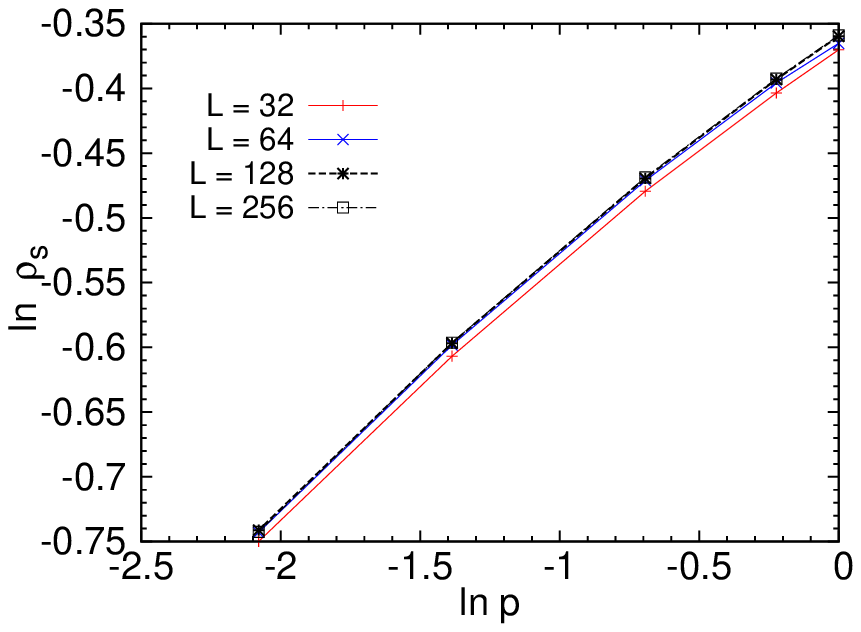}}
  \caption{Variation of $\rho_s$ with  $p$ in log-log
scale for different values of $L$.  \label{por02}}
\end{minipage} %
\begin{minipage}{0.49\linewidth}
\center
  {\includegraphics[width=\linewidth]{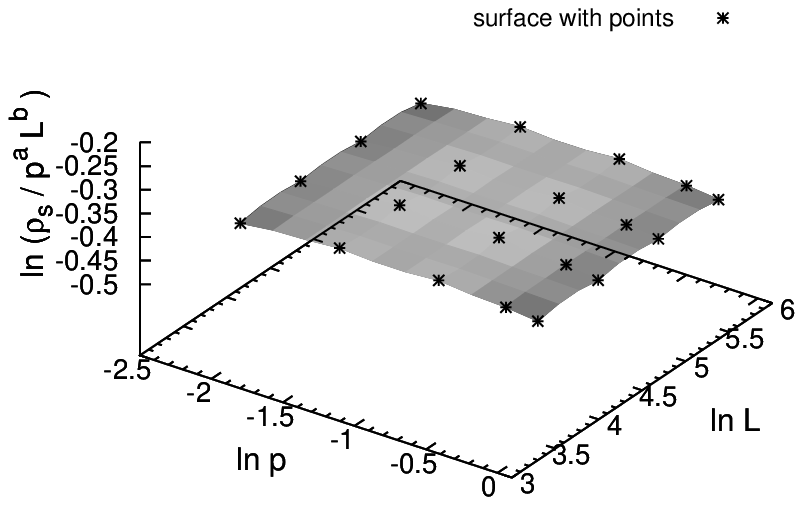}}
  \caption{$\ln \rho_s/p^a L^b$ as a function of $\ln L$ and $\ln p$. \label{porscale}}
\end{minipage} 
\end{figure}

The conductivity $\sigma$
depends on the porosity $\rho$ as well as the geometry of the pore structure.
In order to study the dependence of $\sigma$ on $\rho$, 
an inactive sample of dimensions $L^3$
is considered where no more deposition can take place.
The solid deposited particles are insulating whereas the voids are considered
as conducting when filled with a conducting liquid, such as brine.
Thus, the connected pore space is first identified using Hoshen-Kopelman algorithm 
\cite{hosh76,stau92}. Laplace's equation is solved for the potential distribution
to obtain the conductivity. The discrete version of the Laplace's 
equation $\nabla^2 V(x,y,z) = 0$ is given by
\beq\label{ceq1}
\frac{V_{i-1,j,k} - 2V_{i,j,k} + V_{i+1,j,k}}{(\triangle x)^2} +
\frac{V_{i,j-1,k} - 2V_{i,j,k} + V_{i,j+1,k}}{(\triangle y)^2} 
+ \frac{V_{i,j,k-1} - 2V_{i,j,k} + V_{i,j,k+1}}{(\triangle z)^2} = 0 .
\eeq

\begin{figure}[!htb]
\begin{minipage}{0.49\linewidth}
\center
    {\includegraphics[width=\linewidth]{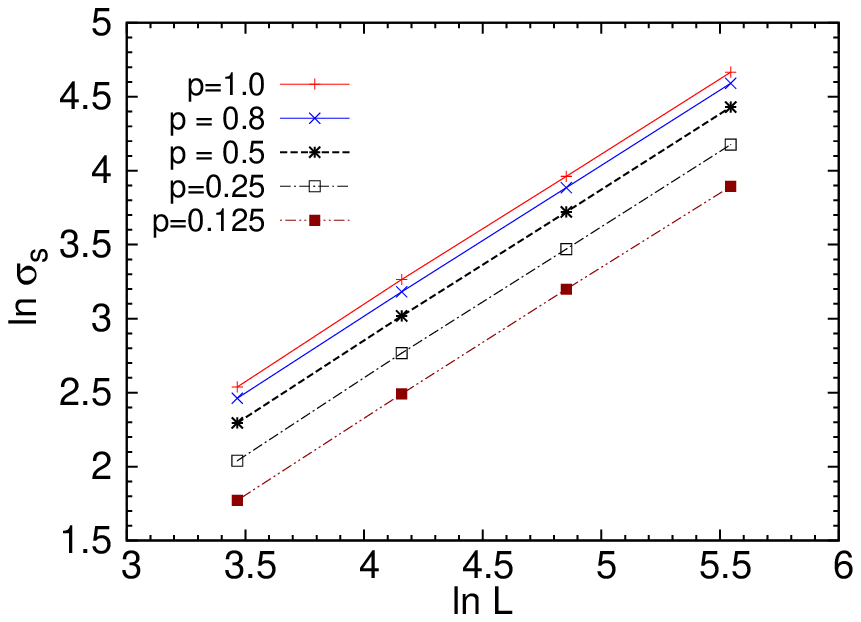}}
    \caption{Variation of $\sigma_s$ with system size $L$ in log-log 
      scale for different values of $p$. \label{cond01}}
\end{minipage} %
\begin{minipage}{0.49\linewidth}
%
\center
    {\includegraphics[width=\linewidth]{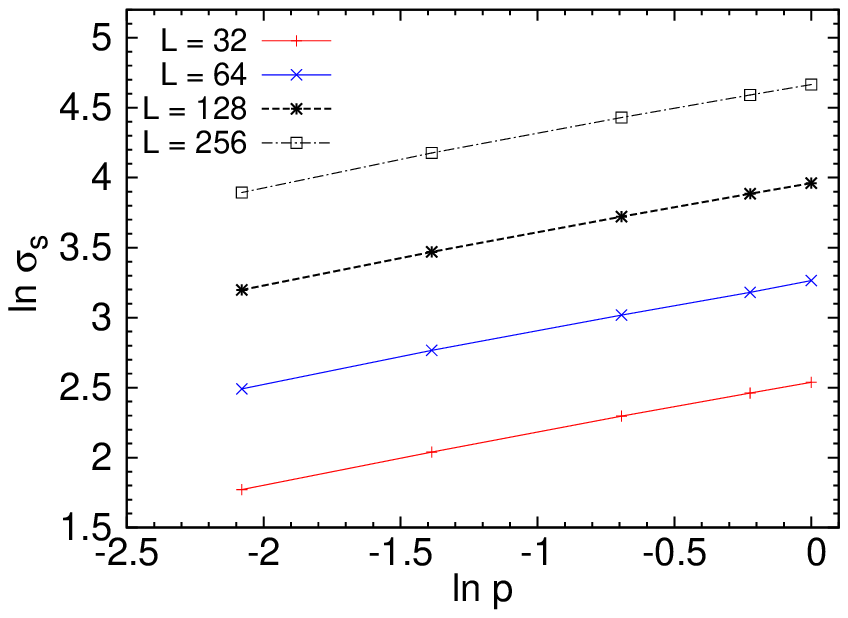}}
    \caption{Variation of $\sigma_s$ with $p$ in log-log
      scale for different values of $L$. \label{cond02}}
\end{minipage} 
\end{figure}

\begin{figure}[!htb]
\center
\begin{minipage}{0.49\linewidth}
  {\includegraphics[width=.9 \linewidth]{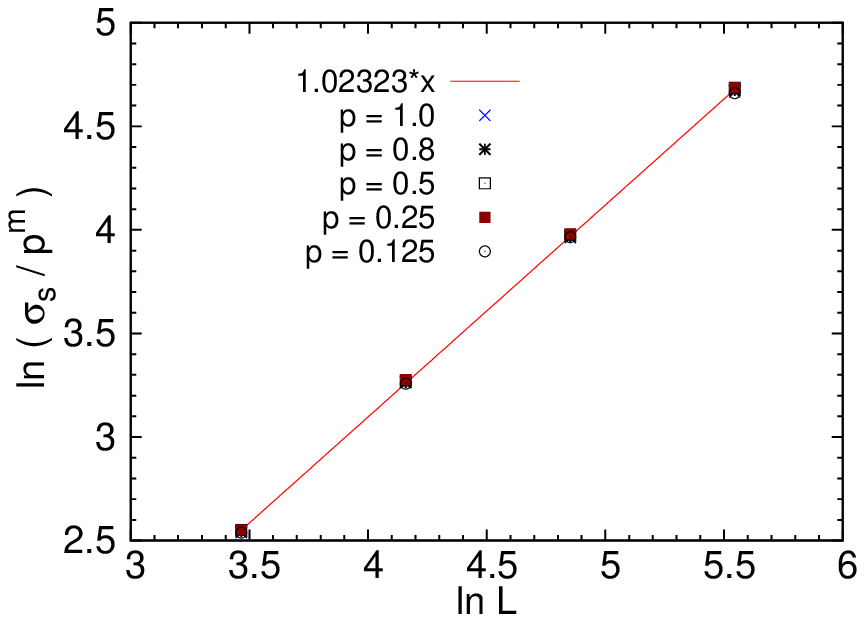}}
  \caption{$\ln \sigma_s /p^m$ versus $\ln L$ \label{cond03}}
\end{minipage} %
\begin{minipage}{0.49\linewidth}
\center
  {\includegraphics[width=1.0\linewidth]{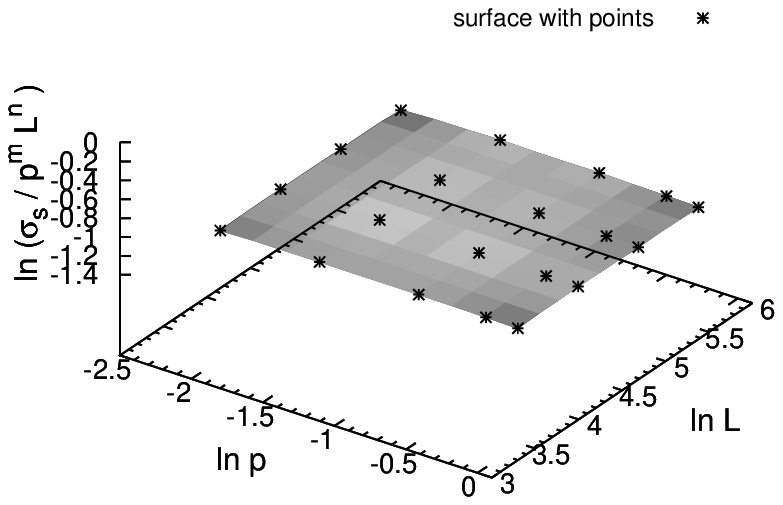}}
  \caption{$\ln \sigma_s/p^m L^n$ as a function of $\ln L$ and $\ln p$ \label{condscale}}
\end{minipage} 
\end{figure}

\begin{figure}[!htb]
\center
  {\includegraphics[width=0.5\linewidth]{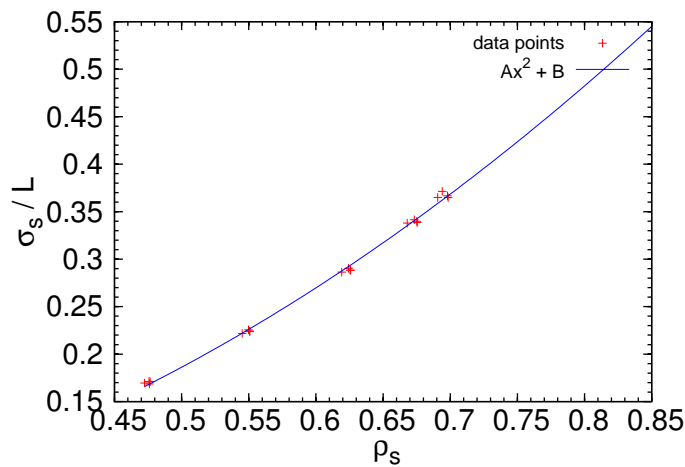}}
  \caption{Variation of $\sigma_s/L$ on $\rho$  \label{condvspor}}
\end{figure}

We consider $(\triangle x) = (\triangle y)= (\triangle z) = 1$, as deposition
consists of unit cubes. Thus, Eq \ref{ceq1} becomes, 
\beq
V^{(n+1)}_{i,j,k} = \frac{1}{N}[V^{(n)}_{i-1,j,k} + V^{(n)}_{i,j-1,k} 
+ V^{(n)}_{i,j,k-1} +V^{(n)}_{i+1,j,k} 
+ V^{(n)}_{i,j+1,k} +V^{(n)}_{i,j,k+1}]
\eeq
where, $N$ is the number of nearest neighbor vacant sites. The boundary conditions
are $V(z=0) = 0, V(z = L) = 1$. The
initial condition obtained by setting values of $V^{(n)}$ for $n = 0$.
We have started with $V^{(n)} = (k/L)$ for all vacant sites
at a height $k$.
Numerically, the steady state is said to be achieved if ($|V^{(n+1)} - V^{(n)}| < \epsilon$), 
where $\epsilon$ is the required accuracy. If the steady state condition
is achieved after $r$ iterations then $V^r$ is the steady state potential.
The conductivity is also found to be a  constant $\sigma_s$ for a fixed
system size $L$ and $p$. 
For a given $L$ and $p$, $\sigma$ reaches a steady state value $\sigma_s$
almost simultaneously with porosity $\rho$.
$\sigma_s$ depends on system size $L$ and sticking probability $p$
as $\sigma_s \sim p^m \, L^n$ with $m=.368496$ and $n=1.02323$
(Fig. \ref{cond01} and Fig. \ref{cond02}). 
Finally, the dependence of $\sigma_s$ on $\rho_s$ is found to satisfy
Archie's law of the form $\sigma_s \sim \rho_s^f$ with $f = 2.02$ as shown
in Fig. \ref {condvspor}.
An excellent collapse is obtained for $\sigma_s$ as shown in Fig. \ref{cond03}
and Fig. \ref{condscale} with $m = 0.368496$ and $n=1.02323$.

\section{Conclusion}
We have studied a generalized ballistic deposition (GBD) model with
deposition of particles having intermediate stickiness 
in (2+1) dimension. Surface width, porosity and conductivity, when such structure is saturated
with conducting liquid, show scaling behaviour with both system
size $L$ and sticking probability $p$.
Scaling relation of surface width is studied and correct
scaling exponents are determined.
The deposition process leads to porous structures.
Scaling of saturated porosity and conductivity with 
system size and sticking probability are also studied
and the corresponding scaling exponents are calculated.
The scaling exponents thus obtained agree with Archie's
law for the dependence of the conductivity on porosity.
A study of generalised deposition with more species of particles having varying intermediate
stickiness $1>p_k>0$, with $k=1,2,3,... \nu$, where $\nu$ is the number of such species, may
be of interest in depository rocks. Such a study is in progress and the results will be reported
elsewhere.
\bibliographystyle{apsrev4-1}
\bibliography{gbd_2d}

\end{document}